\documentclass[final,superscriptaddress,english,twocolumn,amssymb, nobibnotes, aps, PRB,longbibliography]{revtex4-1}

\usepackage{graphicx}
\usepackage{float}

\usepackage{amsmath}
\usepackage{amssymb}
\usepackage{appendix}
\usepackage[dvipsnames]{xcolor}
\usepackage{tikz}
\usepackage{ulem}
\usepackage{placeins}
\usepackage{ifthen}
\usepackage{bm}
\definecolor{darkblue}{rgb}{0,0,.65}
\definecolor{darkgreen}{rgb}{0.28,0.41,0.19}
\usepackage{enumitem}

\usepackage[%
  pdfauthor={Kevin Hemery},%
  pdfstartview=FitH,%
  breaklinks=true,%
  bookmarks=true,%
  colorlinks=true,%
  anchorcolor=black,%
  citecolor=blue,
  filecolor=black,%
  menucolor=black,%
  urlcolor=darkblue,%
  linkcolor=blue,%
 ]{hyperref}
\usepackage[all]{hypcap} 

\definecolor{nicegreen}{RGB}{0,100,0}
\definecolor{coolblack}{rgb}{0.0, 0.18, 0.39}
\newcommand{\kh}[1]{\textcolor{nicegreen} {\textsf{\textbf{KH:}#1 }}}

\usepackage{txfontsb}
\usepackage{times}
\usepackage{mathrsfs}
\usepackage{dsfont}

\begin{document}
	
\title{Identifying Correlation Clusters in Many-Body Localized Systems}%

\author{K{\'e}vin H{\'e}mery}
\affiliation{Department of Physics, TFK, Technische Universit\"at M\"unchen, James-Franck-Stra\ss e 1, 85748 Garching, Germany}
\author{Frank Pollmann}
\affiliation{Department of Physics, TFK, Technische Universit\"at M\"unchen, James-Franck-Stra\ss e 1, 85748 Garching, Germany}
\affiliation{Munich Center for Quantum Science and Technology (MCQST), Schellingstr. 4, D-80799 München}
\author{Adam Smith}
\affiliation{Department of Physics, TFK, Technische Universit\"at M\"unchen, James-Franck-Stra\ss e 1, 85748 Garching, Germany}
\affiliation{School of Physics and Astronomy, University of Nottingham, Nottingham, NG7 2RD, UK}
\affiliation{Centre for the Mathematics and Theoretical Physics of Quantum Non-Equilibrium Systems, University of Nottingham, Nottingham, NG7 2RD, UK}
\email{kevin.hemery@tum.de}

\begin{abstract}
    We introduce techniques for analysing the structure of quantum states of many-body localized (MBL) spin chains by
    identifying correlation clusters from pairwise correlations.
    These techniques proceed by interpreting pairwise correlations in the state as a weighted graph, which we analyse using an established graph theoretic clustering algorithm. 
    We validate our approach by studying the eigenstates of a disordered XXZ spin chain across the MBL to ergodic transition, as well as the non-equilibrium dyanmics in the MBL phase following a global quantum quench. 
    We successfully reproduce theoretical predictions about the MBL transition obtained from renormalization group schemes. 
    Furthermore, we identify a clear signature of many-body dynamics analogous to the logarithmic growth of entanglement. 
    The techniques that we introduce are computationally inexpensive and in combination with matrix product state methods allow for the study of large scale localized systems. 
    Moreover, the correlation functions we use are directly accessible in a range of experimental settings including cold atoms.
\end{abstract}

\maketitle

\section{Introduction}

Initiated by the seminal work of Anderson~\cite{Anderson58}, many-body localization (MBL) is now understood as a dynamical quantum phase of matter ~\cite{Basko2006,Gornyi2005}---defined by the properties of its highly excited many-body eigenstates.
In particular, the entanglement of eigenstates in the MBL phase has been found to obey an area law even at finite energy densities ~\cite{Serbyn2013conservation,Kjall2014,Bauer_2013} and to violate the eigenstate thermalization hypothesis~\cite{Deutsch1991,Srednicki1994}, due to the existence of quasi-local conserved quantities~\cite{Bauer_2013, Serbyn2013,Imbrie2016,Ros2015,Imbrie_Ros_Scardicchio2017}.
The concept of MBL has since proven central to the understanding of several aspects of non-equilibrium physics. 
For instance, MBL is essential to stabilise various emergent Floquet phases of matter, such as discrete time crystals~\cite{Zhang2017,Choi2017}.

The study of MBL has been driven by large scale numerics and  experimental advances in the control of isolated quantum systems.
These efforts have identified characteristic properties of MBL, such as the unbounded logarithmic growth of entanglement following a global quench~\cite{Chiara_2006,Znidaric2008,Bardarson2012,marko2008,Nanduri2014,Serbyn2013Entanglement} --which distinguishes it from Anderson localization (AL) where the entanglement saturates-- and the presence of an eigenstates transition to an ergodic phase at finite disorder strengths~\cite{Luitz2014,PAl2010,Oganesyan2007,Berkelbach2010,Pietracaprina_2017,Geraedts_2017}. 
The logarithmic entanglement growth has since been observed experimentally for small system sizes in Rydberg atomic systems and in superconducting circuits ~\cite{Lukin2019,Chiaro2019}. 
However, extracting the entanglement entropy experimentally generically requires high fidelity measurements of a number of non-local observables that scale exponentially with system size. 
This makes experimental measurements of the entanglement entropy prohibitively difficult for large systems.
In cold atom setups, large systems and long times can be reached, even in 2D, and clear signals of MBL has been detected in local measurements~\cite{Schreiber2015,Choi2016,Bordia2017}. 
In spite of the recent progress, the MBL transition is still not fully understood. 
While we have powerful numerical and analytical techniques that allow us to investigate the slightly entangled eigenstates deep in the MBL phase \cite{Khemani2016, Yu2017, Lim2016, Serbyn2016,Imbrie2016,Ros2015}, the transition to the ergodic phase is much harder to study. 
Phenomenological renormalization group (RG) approaches have emerged as promising theoretical description of the transition~\cite{Potter2015,Morningstar_Huse_RG,Zhang_Huse_RG,RG_Serbyn,Khemani_sparse_thermal_block,Huveneers2017}. Although the assumptions behind the various models differ, most of them describe the MBL transition in term of the proliferation of ``thermal blocks'' versus ``insulating blocks'', i.e., regions of the spin chain that look locally thermal or fully localized, respectively. 
However, the interpretation of these approaches rest on phenomenological assumptions which could bias the results. 
Indeed, most models assume that each of these blocks is local, although the existence of sparse thermal blocks spanning the whole chain has also been suggested \cite{Khemani_sparse_thermal_block}. These RG studies suggest that the MBL transition is of Kosterlitz-Thouless (KT) type with a delocalization mechanism called \textit{avalanche instability}, also sometimes referred as \textit{quantum avalanche}. ~\cite{Morningstar_Huse_RG,Dumitrescu2019}.

Since RG approaches provide a clear mechanism for the transition and allow for a prediction of a scaling behaviour close to the transition, it is desirable to have a clear prescription in order to identify these ``blocks'' from states obtained numerically or experimentally. 
The first numerical validation of this picture in a microscopic model has been provided in Ref.~\cite{multiscale}, where it was proposed to identify these ``blocks'' by finding what they denote as \textit{entanglement clusters}. 
These are clusters of spins that are more strongly entangled with each other than the rest of the system. 
A numerical investigation using exact diagonalization for small systems revealed that the average block size of these entanglement clusters is indeed consistent with the RG analysis of the transition. 
Entanglement entropy is paramount for this approach, but it is costly to calculate both numerically and from experimental measurements.

Motivated by that work, we propose an approach in which we identify these structures in MBL systems in a scalable way that is relevant for efficient matrix product state (MPS) based simulations and accessible in experiments. 
We are focusing on the XXZ spin-chain in the presence of a disordered $z$-directed field defined by the Hamiltonian
\begin{equation}
\hat H= -J \sum_{i} \left[ \hat S^x_{i} \hat S^x_{i+1} + \hat S^{y}_{i} \hat S^y_{i+1} + \Delta \hat S^z_{i} \hat S^z_{i+1} + h_i \hat S^z_{i} \right].
\label{heisenberg}
\end{equation}
The disordered field $h_i$ is sampled uniformly from the interval $[-W,W]$, where $W\geq 0$ controls the strength of the disorder.
We consider the Anderson insulator at $\Delta=0$ as well as the Heisenberg model at $\Delta=1$, which is believed to exhibit an MBL transition at $W_C$ estimated between 2.7 and 3.8~\cite{Luitz2014,PAl2010, Geraedts_2017,Pietracaprina2018, Serbyn2015,De_Luca_2013}.  
%

In this paper, we present practical tools to efficiently identify the ergodic clusters within MBL eigenstates using pairwise correlations by applying methods originally developed in the context of graph theory \cite{invention_modularity,first_girvan_newman,weighted_networks_newman,review_fortunato}.
We validate our approach in two ways:
First, we show that the two site mutual information (TSMI) is a useful proxy for analysing the structure of MBL eigenstates.
Second, we demonstrate in Sec.~\ref{sec:section_dynamics} that our clustering algorithm applied during time evolution--using the TSMI as well as the pairwise connected correlation functions in the $\sigma_z$ basis--indicates the logarithmic spreading of entanglement. 
%

%

%

\section{From correlations to graph theory}
\label{sec:tsmi}
The quantum mutual information of two subsystems $A$ and $B$ is a correlation measure defined as:
\begin{equation}
    I(A;B) = S(A) + S(B) - S(A\cup B),
\end{equation}
where $S(A) = -\text{Tr}[\rho_A \log(\rho_A)]$ is the von Neumann entanglement entropy for the subsystem $A$. The TSMI corresponds to the case where subsystems $A$ and $B$ each consist of a single site, and in this case we denote it $I(i;j)$ for two sites $i$ and $j$. The TSMI captures the classical and quantum correlations between two sites, and has already been shown to be a relevant probe of localization\cite{Giusepe_mutual_info_mbl}. In particular, spatial fluctuations in the TSMI grow logarithmically under non-equilibrium dynamics, mirroring the entanglement entropy~\cite{Giusepe_mutual_info_mbl}. 
Another useful quantity to study quantum correlations is the (connected) correlation function $C(\hat O_A, \hat O_B)$ of two operators $\hat O_A$ and $\hat O_B$:
\begin{equation}
C(\hat O_A, \hat O_B)= \langle \hat O_A \hat O_B \rangle- \langle \hat O_A \rangle \langle \hat O_B \rangle    
\end{equation}
where $\langle \hat O \rangle$ denotes the expectation value of the operator $\hat O$. 
Although TSMI takes into account all pairwise correlations \cite{Wolf2008, Dong_2010}, it is less accessible in experiments than certain correlation functions.
In this paper we introduce tools borrowed from the field of graph theory to extract what we call \textit{correlation clusters}, in analogy to Ref.~\cite{multiscale}. This provides an efficient method for studying of correlations in MBL systems.
Graph theory has been used in the past to detect quantum phase transition in equilibirum settings \cite{Valdez2017,Bagrov2020,Sokolov2020}.  
Recently, another work identified the so called ``ergodic bubbles'' (i.e. regions of space where the expectation values of local operators look thermal) using neural networks techniques \cite{Szoldra2021}.

\begin{figure}[t!]
	\includegraphics[width=\columnwidth]{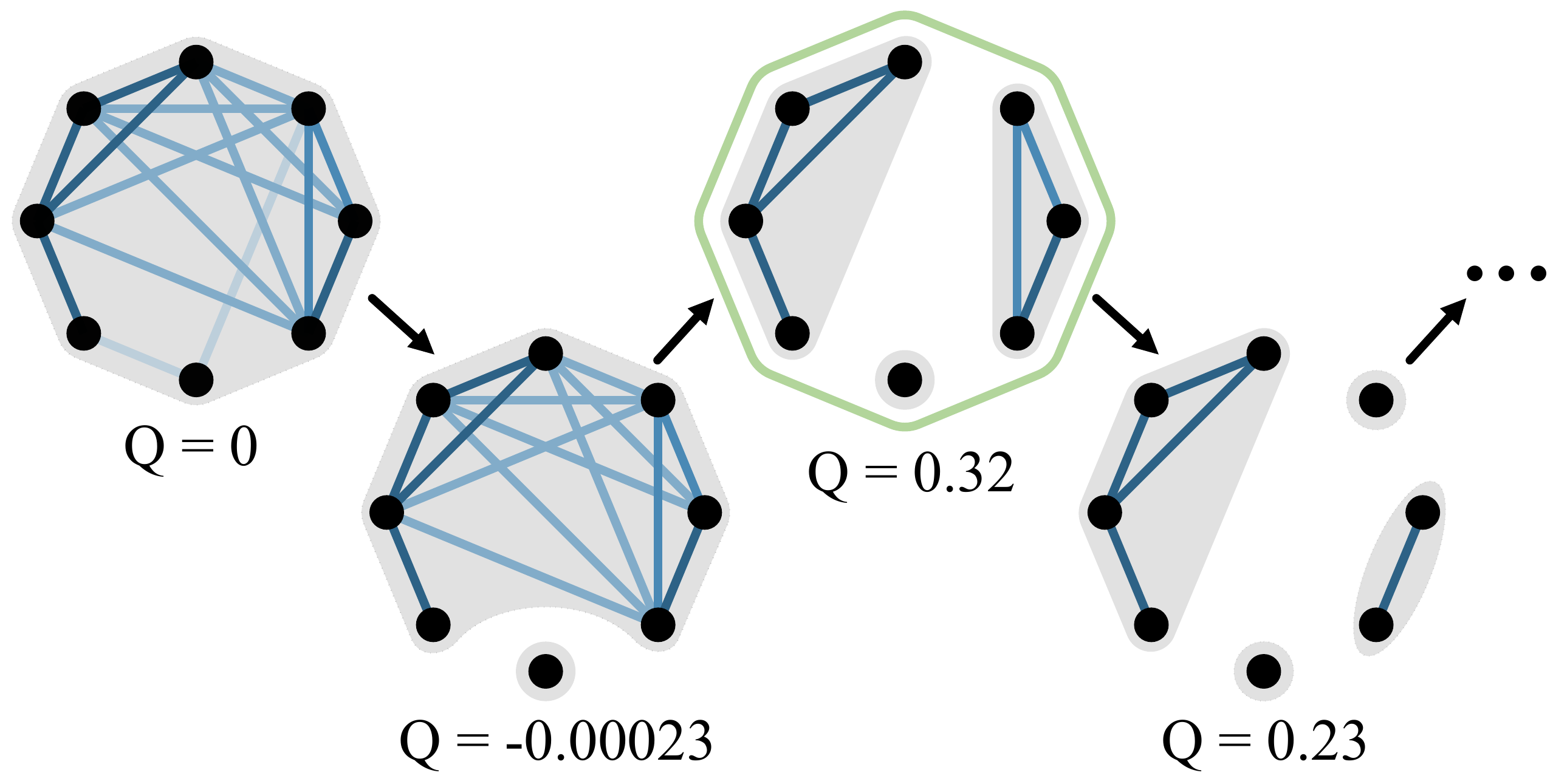}
	\caption{Schematic description of our graph theory approach using an example of the mutual information matrix $M_{ij}$. The full graph has zero modularity (see Eq.~\eqref{Q}). In the second step, the weakest  edges have been successively removed until one site is isolated. This clustering is associated with a very low modularity of $Q=-0.00023$, indicating no community structure. In the third step, the new clustering yields three clusters and a relatively high modularity of 0.32, which turns out to be the highest obtained follwing the procedure. Therefore, we identify this clustering as the physical one. The next steps of the decomposition yields four clusters with a modularity of $Q=0.23$, smaller than in the last step, indicating community structure of a lower quality. The rest of the procedure was not represented here, but the modularity was decreasing at each new step.}
	\label{fig: schematic}
\end{figure}

Our starting point is to construct a matrix $M_{ij}$ containing the correlations between site $i$ and $j$, and to interpret it as an adjacency matrix for a weighted graph as illustrated in Fig.~\ref{fig: schematic}. 
The vertices of this graph are the lattice sites of our system and the bonds connecting them are weighted by the matrix element $M_{ij}$ between that pair. 
Our goal to find the correlation clusters in the state translates to finding ``communities'' within this graph. 
We consider $M_{ij} = I(i;j)$ in the case of eigenstates, to which we add $M_{ij} =C(\hat \sigma^z_i,\hat \sigma^z_j)$ for dynamics.
%

The task of finding communities has received considerable attention in the field of graph theory \cite{invention_modularity,first_girvan_newman,weighted_networks_newman,review_fortunato}. 
This is usually achieved by splitting the graph into disjoint sets of vertices which we refer to as \textit{clusters}. 
A given decomposition of a graph into clusters is referred to as a \textit{clustering}. 
The goal is to find a clustering that is optimal by some well-defined measure. 
Inspired by the well established Girvan-Newman approach \cite{invention_modularity,weighted_networks_newman}, we propose the following three steps procedure, shown schematically in Fig.~\ref{fig: schematic}, for finding the optimal clustering from the correlation matrix $M_{ij}$:
\vspace{5pt}
\setlist{nolistsep}
\begin{enumerate}
\itemsep3pt
\item Successively remove the weakest bonds of the graph. 
\item When the removal of a bond results in two parts of the graph becoming disconnected, we store the new clustering. This clustering corresponds to the set of clusters, where a cluster contains sites that are connected to each other. 
\item Repeating steps 1 and 2 appropriately, we eventually end up with a completely disconnected graph, and have stored a sequence of different clusterings. For each of these stored clustering we then compute the \textit{modularity}:
\begin{equation}
	Q=\frac{1}{2 m} \sum_{ij} \left (M_{ij}-\frac{k_i k_j}{2m} \right) \delta(c_i,c_j),
    \label{Q}
\end{equation} 
where $k_i=\sum_{j}M_{ij}$, and $m=\frac{1}{2} \sum_{ij} M_{ij}$. The delta function $\delta(c_i,c_j)$ is 1 if sites $i$ and $j$ are connected for the given clustering and 0 if they are not. The modularity takes values $Q\in [-1/2,1]$ and quantifies how good the clustering is, with close to 1 corresponding to a good clustering, or ``community structure''. We select the correct clustering as the one with the highest modularity.
\end{enumerate}
The first step differs from the original Girvan-Newman approach. While in our case, we are guided by the physical intuition that two correlation clusters are only connected by weak bonds, Girvan and Newman use a quantity called \textit{edge-betweeness} to assess which bonds are most likely to link separated communities \cite{first_girvan_newman}. 

\section{Correlation clusters in eigenstates}
\label{sec:eigenstates}

\begin{figure}[t]
	\includegraphics[width=\columnwidth]{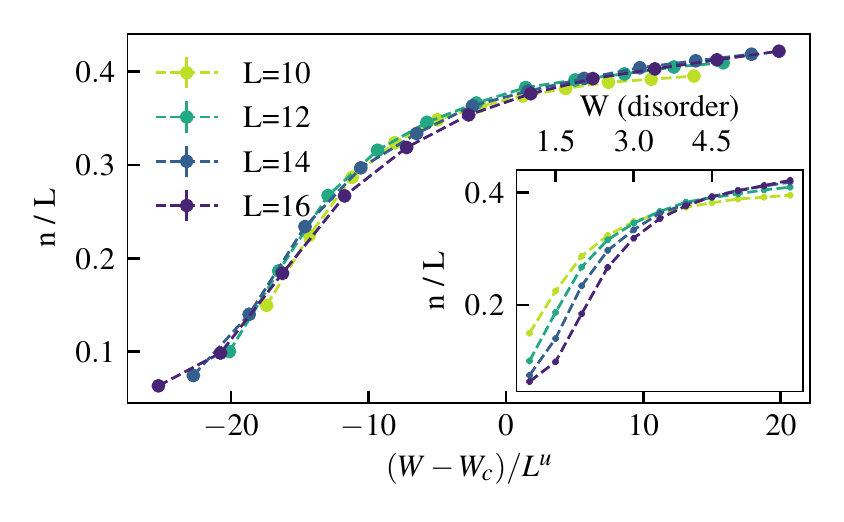}
	\caption{
	 Scaling collapse of the average number of clusters $n$ divided by system size as a function of the disorder obtained using our modified Girvan-Newman approach. For the collapse, the disorder strength was rescaled to take the form $(W-W_c)/L^u$ with $W_c=3.8$ and $u=1.26$. When the modularity was lower than $Q_{\text{th}}=\alpha (1-a/L)$ with $\alpha=0.3$ and $a=3.59$, the state was considered as fully ergodic and made of only one cluster. The coefficient $a$ has been fitted to take into account the finite size effects at $W=6$, according to the finite size scaling: $Q_L=q-\frac{a}{L}$ of the average modularity $Q_L$ obtained for system size L. Inset: average number of clusters $n$ divided by system size as a function of disorder strength $W$.}
	\label{fig:collapse}
\end{figure}

We will now focus on the clustering in mid-spectrum eigenstates for the Hamiltonian Eq.~\eqref{heisenberg} for different values of the disorder strength. 
We analyse the structure of the optimal clustering for the eigenstates across the MBL-ergodic phase transition, using the TSMI to define the graph $M_{ij}$.
It has been shown in earlier studies that the number of entanglement clusters can be taken as a relevant scaling parameters for the MBL transition~\cite{multiscale}. 
In order to validate our graph clustering approach, we perform a similar scaling analysis.
The average number of correlation clusters $n$ as a function of disorder is shown in Fig.~\ref{fig:collapse} for different system sizes. 
We select $50$ eigenstates from the middle of the spectrum of $700$ disorder configurations and then apply the algorithm outlined in the previous section to extract the average number of clusters in the optimal clustering. 
The data collapses convincingly with scaling $n/L=f((W-W_c)/L
^u)$ with parameters $W_c=3.8$ and $u=1.26$, taken from Ref.~\cite{multiscale}. 
It was pointed out that this scaling is consistent with theoretical studies, where a Harris-type bound on the exponents has been derived~\cite{Harris_criterion}.

We note that if the system is ergodic, then we would expect the mutual information to be uniform on average between all pairs of sites \cite{Giusepe_mutual_info_mbl}. 
In this case, the optimal clustering is a single cluster containing all sites but the algorithm as defined will instead choose a clustering with very low modularity. We therefore need to bypass the graph theory algorithm by setting a threshold $Q_{\text{th}}$ below which the states yielding a modularity $Q < Q_{\text{th}}$ are considered as ergodic. 

Since our numerics are performed on finite system sizes up to $L=16$, the modularity will be affected by finite size effects that we must take into account in $Q_\text{th}$. To understand these effects we consider states deep in the MBL phase where we can make statements about the optimal clustering. In particular, MBL eigentates are simultaneous eigenstates of an extensive number of exponentially localized $l$-bits with a characteristic localization length~\cite{Serbyn2013}. This means that the structure of the clustering should be independent of systems size, so long as it is sufficiently large compared to the localization length. As explained in appendix~\ref{sec:scaling modularity}, this actually results in a system size dependence of the modularity for similar clusterings. To account for this we use the system size dependent threshold
\begin{equation}
    \label{eq:scaling_threshold}
    Q_{\text{th}}(L)=\alpha\left(1-\frac{a}{L}\right),
\end{equation}
where $\alpha \in [0,1]$. In practice we obtain the coefficient $a$ by fitting $Q(W=6,L)$, where $W=6$ is the maximum disorder strength considered in our scaling analysis and is located deep within the MBL phase. In the main text, we present results for the overall cutoff parameter $\alpha=0.3$. We show in Appendix \ref{sec:thresholds} that as long as $\alpha$ gives the correct clustering behaviour deep in the MBL and ergodic phases, the scaling collapse is not sensitive to the specific choice of this coefficient.

\begin{figure}[t]
	\includegraphics[width=0.9\columnwidth]{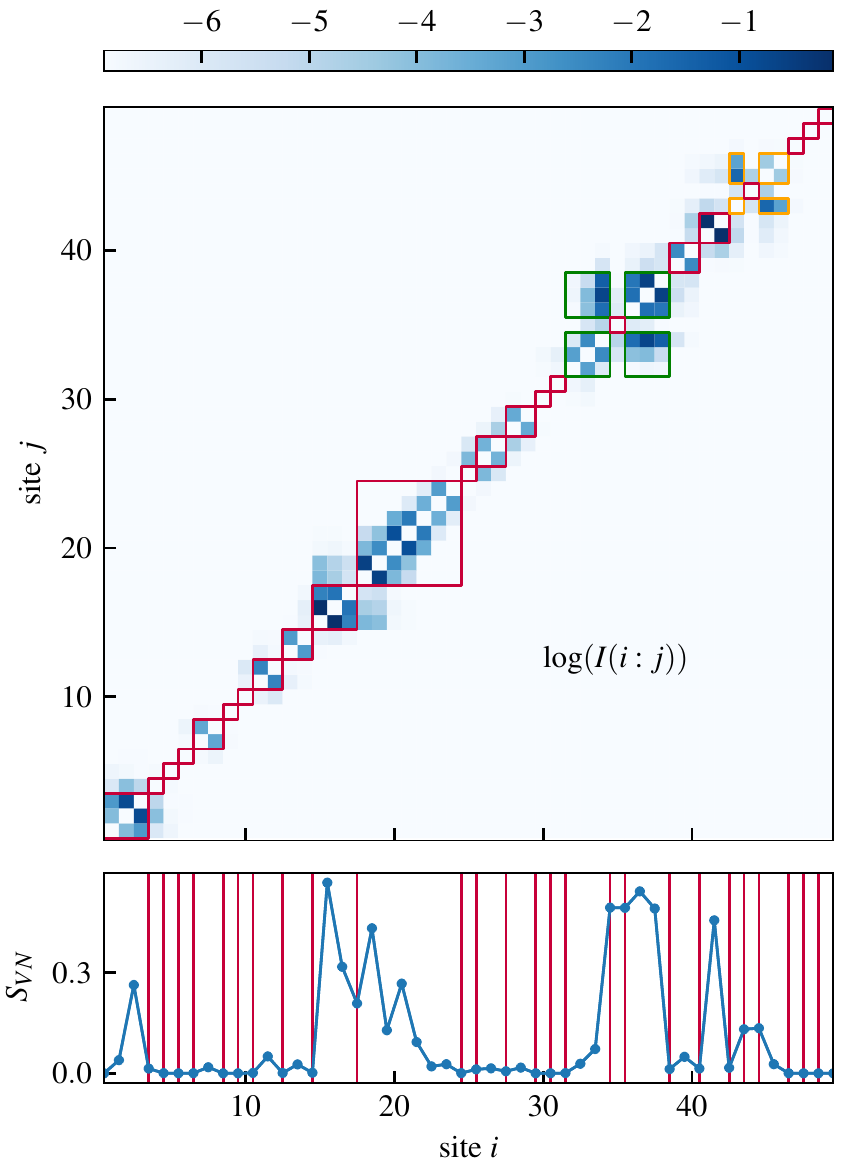}
	\caption{Example of the mutual information matrix and the associated communities (correlation clusters) of an eigenstate of a MBL hamiltonian obtained using the DMRG-X algorithm~\cite{Khemani2016}. This disorder strength is $W=12$, the system size is $L=50$. On the top panel, we plot the mutual information matrix. We draw boxes around the matrix elements belonging to the same ``correlation cluster''. We use a red box when a cluster is connected (i.e. no leapfrogging), while we use orange and green boxes for the two disconnected clusters. On the bottom panel, we present the bipartite entanglement entropy as a function of sites. The boundary between two clusters is signaled by a vertical red line. The boundaries between two disconnected clusters are always associated with a local minimum of entanglement entropy.}
	\label{fig: matrix plot}
\end{figure}

After focussing on the average cluster size, we will now investigate the structure of individual eigenstates using the clustering algorithms. 
The TSMI matrix $M_{ij}$ is shown in Fig.~\ref{fig: matrix plot} for a single mid-spectrum eigenstates in an $L=50$ system with disorder strength $W=12$--obtained using DMRG-X~\cite{Khemani2016}--and compared against the bipartite von Neumann entanglement entropy for cuts along different bonds. 
Here we can see that the localized state is decomposed into a sequence of small clusters (red boxes) and there are only weak off-diagonal (long-range) correlations in the matrix. 
However, we observe several examples of clusters that contain sites that are not nearest neighbours, a phenomenon which, following Ref.~\cite{multiscale}, we refer as ``leapfrogging'' (green and yellow boxes). 
Ideally, we would like to be able to average over many eigenstates obtained by MPS methods on the MBL side of the transition, and to therefore extrapolate its scaling. 
However, given the current state of algorithms, we find this goal impossible to achieve due to the bias in the sampling of the states \cite{simps}.

A few comments are in order:
First, the clustering algorithm is a numerically very inexpensive procedure which is easily scalable, since only two-sites correlations need to be computed, allowing us to apply it to state in the MPS form.

Second, there is a clear agreement between the strong correlations and the increase in entanglement, as it can be seen by comparing the TSMI matrix with the bipartite von Neumann entanglement entropy (see Fig.~\ref{fig: matrix plot}). 
Indeed, the separation between two local communities always coincide with a local minimum of bipartite entanglement entropy. 
However, our approach gives information that could not be directly extracted from the bipartite entanglement entropy. 
In particular, not all local minima of entanglement entropy signal a separation between two clusters. 
For this reason, the bipartite entanglement entropy alone is not a good criterion to detect clusters since one would need to arbitrarily fix a cutoff entanglement entropy to determine where the separation between two communities is. 
Moreover, entanglement entropy is unable to detect non-local clusters, i.e.``leapfrogging'', which we detect with our graph theory approach.

This brings us to our third point, namely that our approach does not rest on \textit{a priori} physical assumptions, such as locality of the clusters for example.
%
Indeed, the graph theory algorithm does not know about the spatial arrangements of the sites, since its only input is the TSMI matrix.
%
%
However we note that in all cases we considered, the clusters were still relatively local and did not extend throughout the system, in accordance with the results of Ref. \cite{multiscale}. 

\section{Non-equilibrium dynamics}
\label{sec:section_dynamics}
We now turn to the behavior under non-equilibrium dynamics in the localized phase and compare AL and MBL systems. 
We now consider a global quantum quench protocol, starting from an initial N{\'e}el state $|\cdots \uparrow\downarrow\uparrow\downarrow \cdots \rangle$, and time evolve using the Hamiltonian Eq.~\eqref{heisenberg} with $\Delta = 1$ (MBL) or $\Delta = 0$ (AL). 
We can then analyse the correlations as a function of time and identify the time dependence of the correlation clusters. We compare results obtained using the TSMI, $M_{ij}=I(i:j)$, and the correlation functions, $M_{ij}=C(\hat \sigma_i^z,\hat \sigma_j^z)$.

\begin{figure}
	\includegraphics[width=\columnwidth]{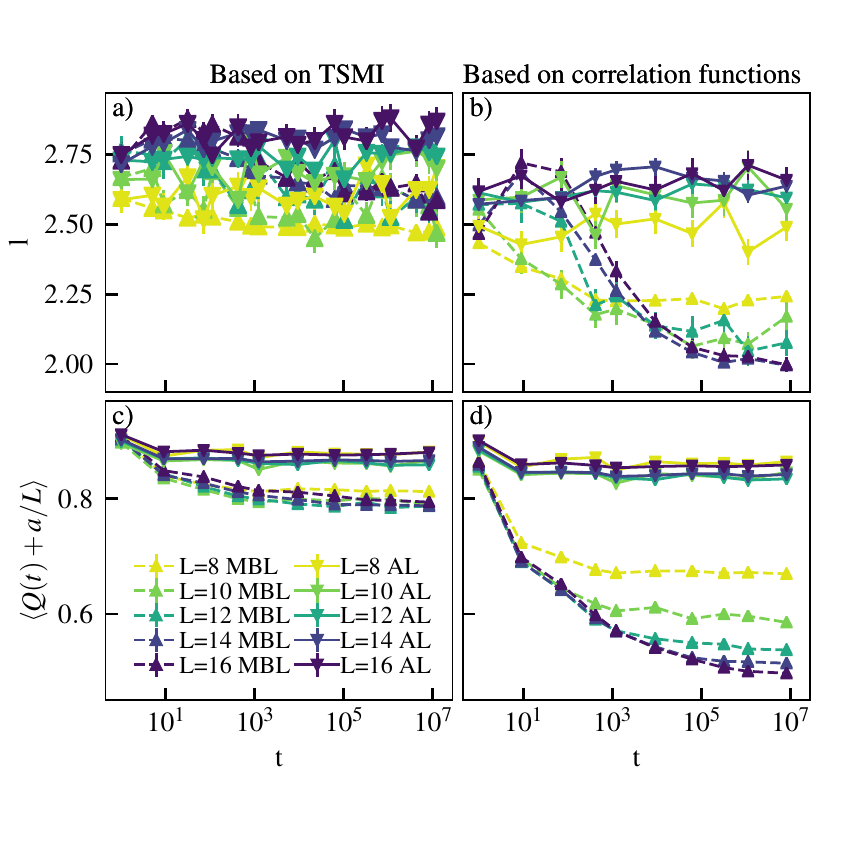}
	\caption{Average length of the clusters (panels a) and b)) and average modularity (panels c) and d)) as a function of time for different system sizes, with disorder strength $W=8$, for both an Anderson localized (AL) ($J_z=0$) and MBL Hamilonian ($J_z=1$). We start from a N\'eel state, simulate a quench using exact time evolution (ETE) and apply our graph theory approach to the TSMI matrix (panels a) and c))  and to the pairwise correlation functions in the $\sigma_z$ basis (panels b) and d)). The fitting parameters are the following: panel c) MBL: $a=3.69$, AL: $a=3.63$; panel d) MBL: $a=3.76$, AL: $a=3.77$.}
	\label{fig: l time}
\end{figure}

Fig.~\ref{fig: l time} a and b show the numerical results for the average cluster length $l$ as a function of time.
When using $M_{ij}=I(i:j)$, $l$ stays approximately constant throughout time, both in the interacting and non interacting cases. 
In contrast, when using $M_{ij}=C(\hat \sigma_i^z,\hat \sigma_j^z)$, $l$ decreases in the MBL case while it stays constant in the AL case. 

In order to understand these results better and to be  able to distinguish further MBL from AL using graph theory, we show the numerical results for the average modularity as a function of time on Fig.~\ref{fig: l time} panels b) and d).
The offset of the modularity has been shifted so that the values for different system sizes coincide at short times. 
Indeed it is shown in Appendix \ref{sec:scaling modularity} that the modularity scales with the system size  as $Q \sim q- a L^{-1}$ for comparable clusters. 
The value of $a$ is found by fitting the data at short times and we find it to be roughly the same for both AL and MBL.
In the non interacting case, the modularity $Q$ stays constant throughout the time evolution. 
On the contrary, $Q$ decreases in the interacting case.

These observations can be explained as follows: at very short times (of the order of $\frac{1}{J}$), correlation clusters appear similarly for both AL and MBL. 
Due to dephasing in the MBL case, these clusters interact exponentially slowly with separation between them, leading to a logarithmically slow decrease of the modularity until it reaches a minimum set by the system size.
These additional correlations over longer ranges can lead to the clusters breaking down further. 
Nonetheless, due to the presence of l-bits in the MBL system, these clusters are robust. 

When using the TSMI as adjency matrix, the clustering overall stays the same and the dephasing can be seen only through the modularity which signal the interaction between the clusters. 
When using the correlation functions, the picture stays roughly the same, although the average cluster size decreases slightly. 

The difference of behaviour in the cases of TSMI and correlation functions stems from the lack of transport in MBL ~\cite{Luitz2020}, which implies that off-diagonal correlation functions cannot build up beyond the localization length. Thus, for our charge conserving model, only the $\sigma^z$ component contributes to the growth of the TSMI at long times. 
Numerical evidence for the spreading of correlation functions in the $\sigma_z$ basis is presented  in appendix ~\ref{sec: propagation_z}.
As a consequence, when using the TSMI, the information contained in the $\sigma^z$ correlation functions is washed out by all the other correlations, which necessarily decay for sufficiently large distances. This leads to more robust clusters which interact less strongly with each other.
This is in line with findings of previous works ~\cite{Smith2016,Tomasi19,Safavi_Naini2019}, which have shown that quantities based on these correlators, in particular certain types quantum Fisher information, can probe the logarithmic growth of entanglement in MBL systems. 

These findings show that it is advantageous to consider the $\sigma^z$ component in this context. In particular, the diagonal $\sigma^z$ correlations are accessible in existing quantum gas microscope experiments \cite{Schreiber2015, Bakr2009, Sherson2010} and thus our technique can be directly be applied in such settings. 

%

\section{Conclusion}
\label{sec:discussion}

In this work, we have shown how to efficiently investigate the structure of MBL states using pairwise correlation functions and TSMI. 
We focused on two applications. First, we provide new numerical techniques for probing the structure of MBL eigenstates, scalable to large systems, particularly relevant for states obtained by MPS methods.  
Second, we show that our approach can provide a characterization of dynamics in the MBL phase.
We have demonstrated that our clustering procedure yields results physically consistent with previously known results. 
When looking at the eigenstates, the scaling of the length of the clusters found in previous works~\cite{multiscale} has been recovered. When looking at the dynamics, our results were consistent with the dephasing process between distant l-bits which is observed in other quantities such as the entanglement entropy or the quantum Fisher information. 
Moreover, we found that the quality of the clustering at late times was still high, a fact which underlines the relevance of the clustering in the time evolution of MBL systems, for it is the persistence of these relatively well separated clusters which prevents full thermalization of the state and keeps the saturation of entanglement entropy well below the Page value. 

More broadly, we have demonstrated the possibility of probing the structure of quantum states based solely on pairwise correlations using graph theory approaches.
\begin{acknowledgments}
    We would like the thank Giuseppe de Tomasi for the stimulating discussions. This work was supported by the European Research Council (ERC) under the European Unions Horizon 2020 research and innovation program (grant agreement No. 771537). A.S. was supported by a Research Fellowship from the Royal Commission for the Exhibition of 1851. F.P. acknowledges the support of the Forschungsgemeinschaft (DFG, German Research Foundation) under Germany's Excellence Strategy EXC-2111-390814868 and the DFG TRR80.
\end{acknowledgments}
\FloatBarrier

\bibliography{bibliography}
\FloatBarrier

\appendix
\section{Scaling with different modularity thresholds}
\label{sec:thresholds}
\begin{figure}
\label{fig:coeff025}
    \centering
    \includegraphics{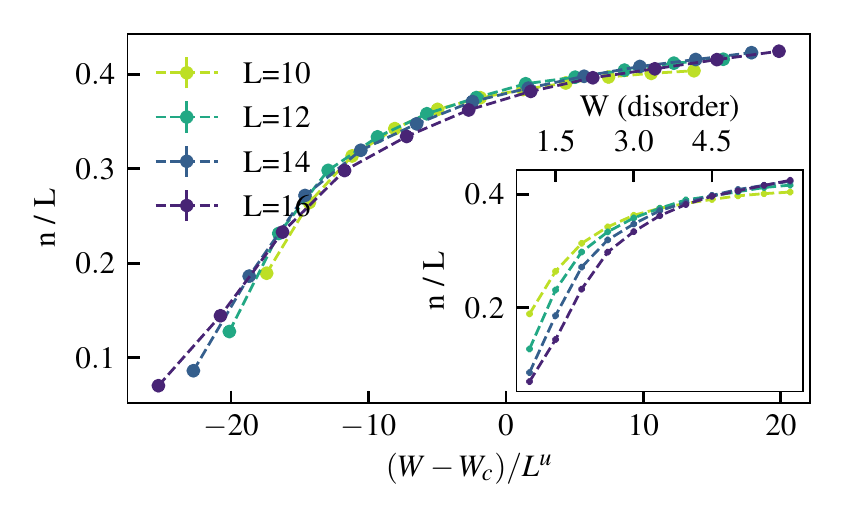}
    \caption{Scaling collapse of the average number of clusters $n$ divided by system size as a function of the disorder obtained using our modified Girvan-Newman approach and $\alpha=0.25$. The parameters used for the scaling collapse are the same as in the main text. Inset: average number of clusters $n$ divided by system size as a function of disorder strength $W$.}
    
\end{figure}

\begin{figure}
    \label{fig:coeff04}
    \centering
    \includegraphics{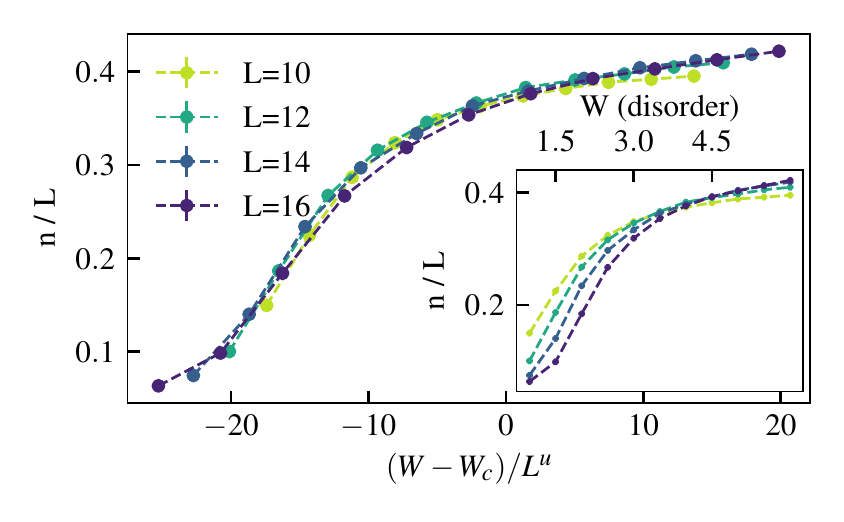}
    \caption{Scaling collapse of the average number of clusters $n$ divided by system size as a function of the disorder obtained using our modified Girvan-Newman approach and $\alpha=0.4$. The parameters used for the scaling collapse are the same as in the main text. Inset: average number of clusters $n$ divided by system size as a function of disorder strength $W$.}
    
\end{figure}

In the main text, we present the scaling collapse of the number of clusters divided by system size. When the modularity obtained for one clustering is smaller than $Q_{\text{th}}(L)=\alpha (1-a/L)$, we bypass our algorithm and consider that the state is fully ergodic and therefore made of a single cluster. In Figs. \ref{fig:coeff025} and \ref{fig:coeff04}, we show that the scaling collapse is not sensitive to the value of the coefficient $\alpha$, as long as $\alpha$ is such that the modularity of almost all eigenstates deep in the ergodic (resp. MBL) phase is below (resp. above) $Q_{\text{th}}(L)$.


\section{Scaling collapse of the number of clusters using the pairwise correlation functions in the $\sigma^z$ basis}
\label{sec:zz_collapse}

\begin{figure}
    \centering
    \includegraphics{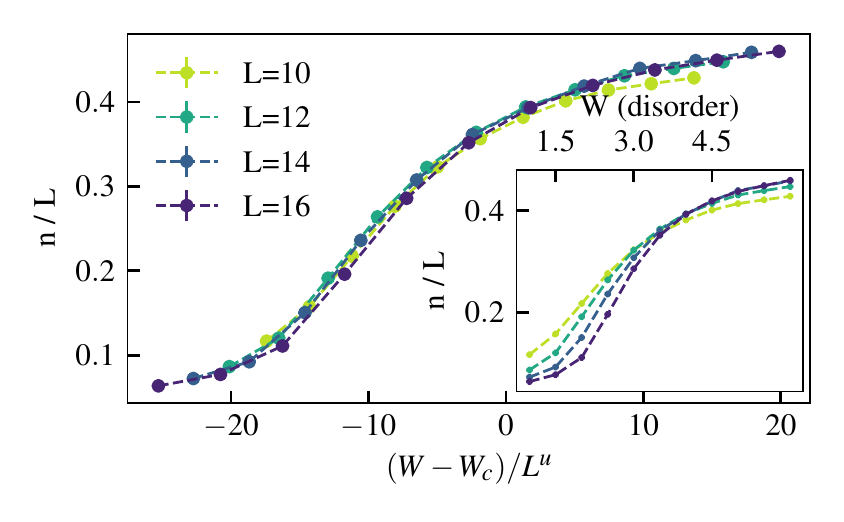}
    \caption{Scaling collapse of the average number of clusters $n$ divided by system size as a function of the disorder obtained using our modified Girvan-Newman approach applied to the pairwise correlation functions in the $\sigma^z$ basis, $\alpha=0.15$ and $a=3.01$. The parameters used for the scaling collapse are the same as in the main text. Inset: average number of clusters $n$ divided by system size as a function of disorder strength $W$.}
    \label{fig:zz_collapse}
\end{figure}

In the main text, we present in Fig.~\ref{fig:collapse} a scaling collapse of the averaged number of clusters divided by system size for which we used the TSMI of the eigenstates as the adjency matrix in our graph theory approach. We show in Fig.~\ref{fig:zz_collapse} that the same approach using the pairwise correlation functions in the $\sigma^z$ basis yields the same scaling collapse. In order to ensure that all states deep in the MBL phase are identified as such, we need to choose a smaller coefficient $\alpha$ for $Q_{\text{th}}$ (see Eq. ~\eqref{eq:scaling_threshold}). Here we choose $\alpha=0.15$.

\section{Details about the modularity}
We would like to have a quantity enabling us to judge the quality of a graph clustering in order to compare different clusterings. In order to achieve this, a naive trial would be evaluate the fraction of weighted edges connecting vertices belonging to the same community over the sum of all weights. If M is the adjency matrix, and $c_i$ the community in which vertex $i$ belongs to, this ratio can be expressed as:
\begin{equation}
 R=\frac{1}{2 m} \sum_{ij}  M_{ij}\delta(c_i,c_j) 
\end{equation}
where $m=\frac{1}{2} \sum_{ij} M_{ij}$ and $\delta(c_i,c_j)$ ensures that vertex $i$ and $j$ are in the same community. 
  Although any good clustering of a given graph should yield a high value of $R$, this quantity can not be useful to gather information about the community structure. Indeed considering only one community containing all the vertices would result in a maximal value of $R=1$ \cite{invention_modularity}.
  In order to overcome this limitation, the formula of $R$ has been modified by following the idea that a random graph is not expected to present a community structure. Therefore, a good measure of quality would be obtained when comparing the actual density of weight to the one we would have if the edges would have been linked regardless of any community structure. This translate into the following expression for the \textit{modularity} \cite{invention_modularity,weighted_networks_newman,review_fortunato} of a graph partition:
  \begin{equation}
  Q=\frac{1}{2 m} \sum_{ij} \left (M_{ij}-P_{ij} \right) \delta(c_i,c_j)
  \end{equation} 
  where $P_{ij}$ is the expected adjency matrix of the random graph which shares the same structural properties as the original graph of interest without presenting the same community structure. This random graph is also sometimes called the ``Null model''. In order to determine the matrix $P$, we must first specify a choice for the null model. Since it has to be similar to the original graph, we impose that the vertex  of the random graph has to have the same \textit{degrees} $k_i$ than the original one, that is to say:
  \begin{equation}
  \sum_{j}M_{ij}=\sum_{j}P_{ij}=k_i.
  \end{equation}   
In other words, every vertex of the null model shares as much weight with the rest of the system than the graph of interest, although the connections between vertices are assigned randomly. On average, the vertex i and j will be connected by a edge of weight $P_{ij}=\frac{k_i k_j}{2m}$ \cite{review_fortunato}, yielding \cite{weighted_networks_newman}:
\begin{equation}
Q=\frac{1}{2 m} \sum_{ij} \left (M_{ij}-\frac{k_i k_j}{2m} \right) \delta(c_i,c_j).
\end{equation} 
We can see that this measure solves the issue initially encountered since the partition containing all vertices have a zero modularity.
A value of modularity close to zero means that the partition is not better than a random one while a value close to one indicate a strong community structure.
\section{Example of the evolution of the clustering in the dynamical case}
\begin{figure*}
    \centering
    \includegraphics{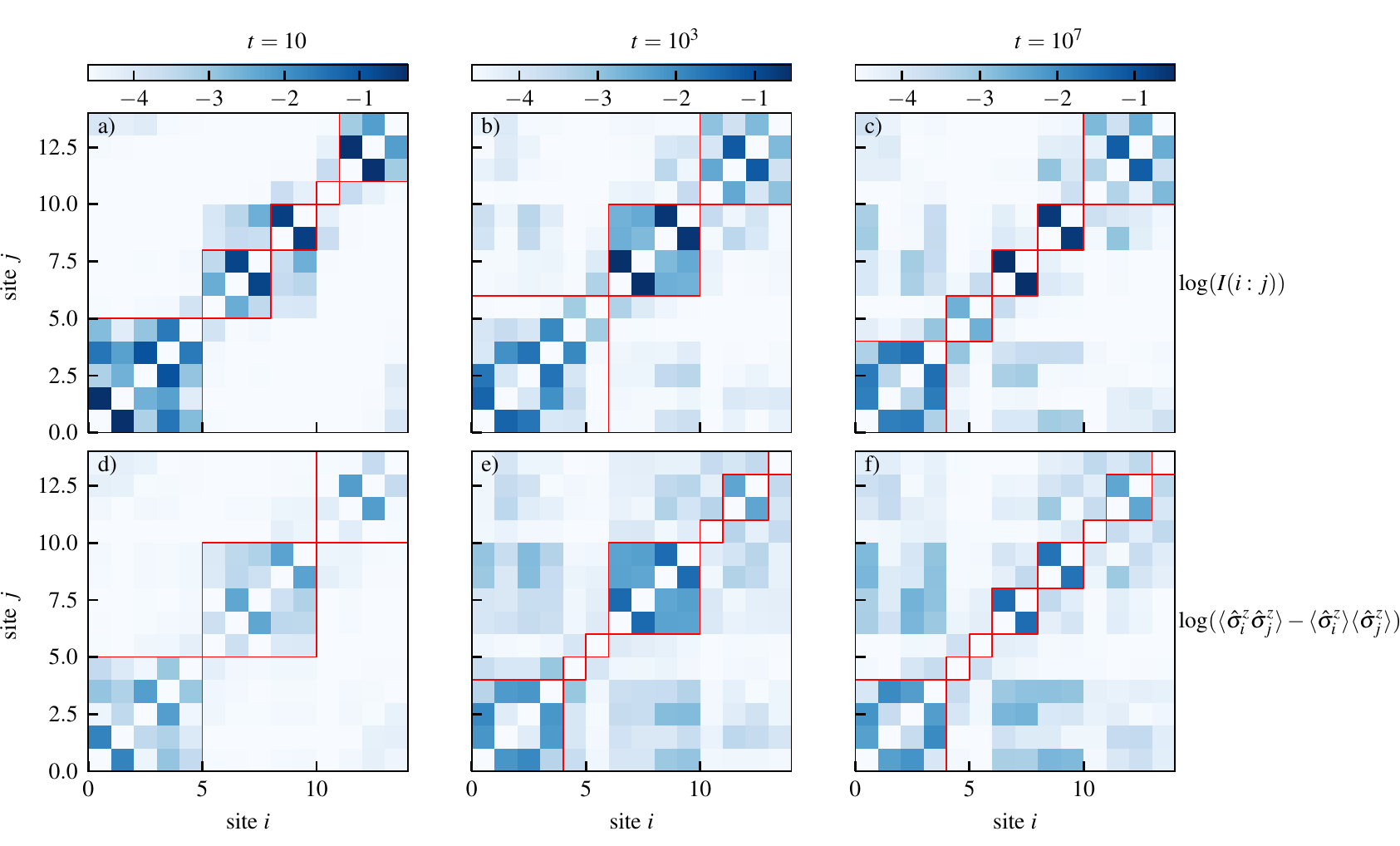}
    \caption{Cluster decomposition for different times obtained for an initial Neel state, time evolved with the Hamiltonian \eqref{heisenberg} with periodic boundary conditions for disorder strength $W=8$ and $L=14$. Panels a and d $t=10$; panels b and e $t=10^3$; panels c and f $t=10^7$. Panels a, b and c: $M_{ij}=I(i:j)$. Panels d, e and f: $M_{ij}=C(\hat \sigma_i^z, \hat \sigma_j^z)$.}
    \label{fig:cluster_t}
\end{figure*}
Fig. \ref{fig:cluster_t} shows the evolution of the clustering for $t=10$, $t=10^3$ and $t=10^7$. At short times, correlations start to build up locally, resulting in the formation of three large clusters. At intermediary times these blocks start to break up as correlation become more non local. Inter cluster correlations (corresponding to ``off-diagonal'' elements on the correlation matrix) are more important resulting in a decrease of modularity. At long times, this process continues to unfold, with a further fragmentation of the cluster structure. However we note that, despite longer range correlations, a clear cluster structure is present, and the correlations are not completely scrambled. Moreover, the inter-cluster interactions is more pronounced in the case of the correlation functions than in the case of the TSMI.

\section{Scaling of the modularity with system size}
\label{sec:scaling modularity}

For a system where the optimal decomposition yields $N$ clusters, the modularity can be written in the following way \cite{Good2010}:
\begin{equation}
Q  =\sum_{i=1}^{N} \frac{e_i}{m}-\left(\frac{d_i}{2m} \right)^2\\
\end{equation}
where the sum runs over the clusters. In the formula above, $d_i$ denotes the total degree of nodes in the cluster $i$: $d_i=\sum_{j} k_j \delta(c_j,c_i)$ in the notation of the main text, $e_i$ is the number of edges in cluster $i$ and m is, as in the main text, the total number of edges. Defining $\langle e \rangle=\frac{1}{N}\sum_{i} e_i$ and $\langle d \rangle=\frac{1}{N}\sum_{i} d_i$ we obtain:
\begin{equation}
Q=\sum_{i=1}^{N} \frac{\langle e \rangle}{m}-\frac{\langle d^2 \rangle}{(2m)^2}    
\end{equation}
We now introduce the quantity $\langle e^{out} \rangle$, which is the average weight going out of each cluster: 
\begin{equation}
\langle e^{out} \rangle= \frac{1}{N} \sum_{i} \sum_{j} M_{i,j} (1-\delta(c_i,c_j)).
\end{equation}
Using the fact that $\langle d \rangle=\langle 2 e  \rangle+ \langle e^{out} \rangle$ and $m=\frac{1}{2}N \langle d \rangle$, we obtain:
\begin{equation}
Q=\sum_{i=1}^{N} \frac{\langle e \rangle}{\frac{N}{2}(2 \langle e \rangle + \langle e^{out} \rangle)}- \frac{\langle (2 e + e^{out})^2 \rangle}{N^2(2 \langle e \rangle + \langle e^{out}\rangle )^2}
\end{equation}

Finally, noting that the number of clusters $N$ is proportional to the system size, we recover the scaling of the main text:
\begin{equation}
    Q=\frac{\langle e \rangle}{\langle e \rangle + \langle e^{out} \rangle/2}-\frac{1}{N} \frac{\langle(2e+e^{out})^2\rangle}{(2 \langle e \rangle+\langle e^{out} \rangle)^2}
\end{equation}
\section{Information propagation using pairwise correlation in the $\sigma_z$ basis}
\label{sec: propagation_z}
\begin{figure}
    \centering
    \includegraphics{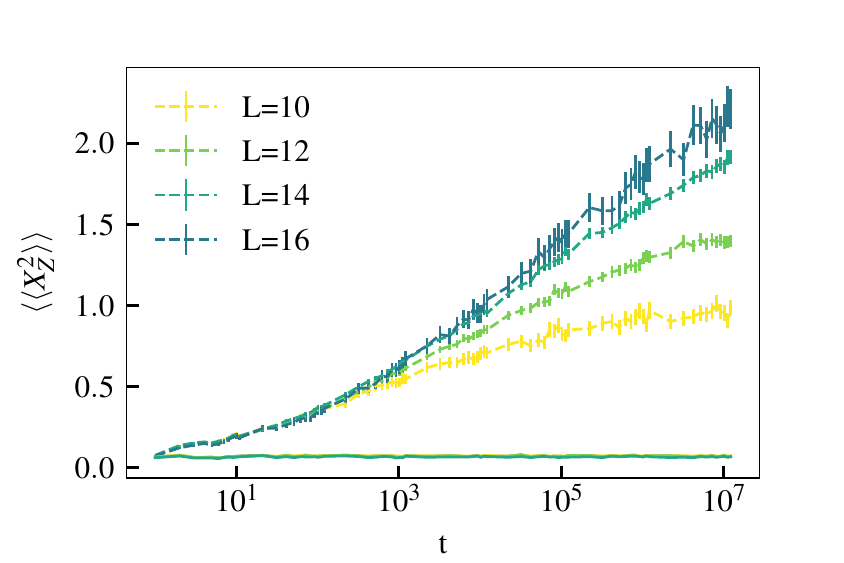}
    \caption{$\langle \langle X_Z \rangle \rangle$ for $W=8$ for various system sizes. The MBL case is shown in dashed line while the AL case is plotted in full lines. All the curves for the AL case are superimposed. This demonstrate that the pairwise correlations in the $\sigma_z$ basis are sufficient to probe the logarithmic propagation of information.}
    \label{fig:variance}
\end{figure}
In Ref.~ \cite{Giusepe_mutual_info_mbl}, it was shown that one can use the TSMI to detect the MBL phase. More precisely, one has to monitor the following quantity during a global quench:
\begin{equation}
    \langle \langle X_I^2 \rangle \rangle = \sum_{j} j^2 I_j(t) - \left ( \sum_{j} j I_j(t) \right)^2
\end{equation}
where $I_j(t)=I(0;j)(t)$. The MBL phase is characterized by a logarithmic growth of $\langle \langle X_I^2 \rangle \rangle$, since this quantity measures the spreading of information in the system. This is explained by the fact that two separate portions of the system need a time exponential with their distance to get entangled. 

 To demonstrate this, we perform ETE with the Hamiltonian \eqref{heisenberg} with open boundary conditions and calculate the following disorder averaged quantity: 
 \begin{equation}
 \langle \langle \hat X_Z^2 \rangle \rangle=\sum_{j} j^2 C(\hat \sigma_0^z, \hat \sigma_j^z,t)- \left ( \sum_{j} j C(\hat \sigma_0^z, \hat \sigma_j^z,t)  \right)^2. 
\end{equation}
This quantity also exhibits logarithmic growth.
\end{document}